\documentclass[sigconf, authordraft=false, anonymous=false]{acmart}

\pdfoutput=1

\AtBeginDocument{%
  \providecommand\BibTeX{{%
    \normalfont B\kern-0.5em{\scshape i\kern-0.25em b}\kern-0.8em\TeX}}}

\renewcommand\footnotetextcopyrightpermission[1]{} 
\usepackage{booktabs} 
\usepackage{graphicx}
\usepackage{balance}
\usepackage{comment}
\usepackage{graphicx,epstopdf}

\usepackage{epsfig}
\usepackage{subfigure}
\usepackage{color}
\usepackage{xcolor}
\usepackage{algorithmic}
\usepackage{algorithm}
\usepackage{eurosym}
\usepackage{balance}
\usepackage{bbm}
\usepackage{tabularx}
\usepackage{textcomp}
\usepackage{gensymb}
\usepackage{multirow}

\acmConference{}{}{}

\begin{document}

\title[eWake]{eWake: A Novel Architecture for Semi-Active Wake-Up Radios Attaining Ultra-High Sensitivity at Extremely-Low Consumption}


\author{Giannis Kazdaridis}
\email{iokazdarid@uth.gr}
\affiliation{%
  \institution{Department of Electrical and Computer Engineering, University of Thessaly, Greece}
}

\author{Nikos Sidiropoulos}
\email{nsidirop@uth.gr}
\affiliation{%
  \institution{Department of Electrical and Computer Engineering, University of Thessaly, Greece}
}

\author{Ioannis Zografopoulos}
\email{zografop@uth.gr}
\affiliation{%
  \institution{Department of Electrical and Computer Engineering, University of Thessaly, Greece}
}


\author{Thanasis Korakis}
\email{korakis@uth.gr}
\affiliation{%
  \institution{Department of Electrical and Computer Engineering, University of Thessaly, Greece}
}

\renewcommand{\shortauthors}{G. Kazdaridis, et al.}

\begin{abstract}
In this work we propose a new scheme for semi-passive \emph{Wake-Up Receiver} circuits that exhibits remarkable sensitivity beyond \emph{-70 dBm},
while state-of-the-art receivers illustrate sensitivity of up to \emph{-55 dBm}.
The receiver employs the typical principle of an envelope detector that harvests RF energy from its antenna,
while it employs a nano-power operation amplifier to intensify the obtained signal prior to 
the final decoding that is realized with the aid of a comparator circuit.
It operates at the 868 MHz ISM band using \emph{OOK signals} propagated through LoRa transceivers,
while also supporting addressing capabilities in order to awake only the specified network's nodes.
The power expenditure of the developed receiver is as low as \emph{580 nA}, remaining at the same power consumption levels as the state-of-the-art implementations.


\end{abstract}




\maketitle


\vspace{-0.2cm}
\section{Introduction}
  \label{intro}
\vspace{-0.1cm}



Energy efficiency is a leading topic of research in the domain of \emph{Wireless Sensor Networks} (\emph{WSNs}).
In most real-world applications, sensor nodes are battery operated, while their life duration is solely dependent on the battery's remaining charge
and the node's power profile.
A common strategy for saving energy in sensor networks is the \emph{duty-cycle} practice, which suggests that sensor nodes
enter a low-power mode, the so-called sleep state, in order to save as much energy as possible during their inactive periods. 
The sleep state is interrupted by short, burst events, where sensors sense, process and propagate data.
The above principle is usually realized using internal or external time keeping circuits that provide fixed interrupt signals to awake the devices from their sleep state.
Moreover, it is common that the interval of the wake-up signals is fixed and predefined depending on the application scenario of the network.
Despite the fact that the aforementioned principle significantly reduces the overhearing and idle listening problem, which is a major source of energy wastage \cite{overhearing}, it is not considered to be the best practice, 
especially in application scenarios that do not require fixed time intervals.

Another method that eliminates the duty-cycle obstacle is the employment of a \emph{Wake-Up Receiver} (\emph{WuR}), presented in a few research works \cite{magno, Mango_journal, wur_Spenza, marinkovic, Magno_LorA_Journal, Gamm, wur_demo, wur_stages}. 
This receiver is actually an auxiliary circuit usually attached to the main sensing device 
in order to notify the latter to switch from its sleep to its active phase.
This circuit typically draws less than \emph{1 $\mu$A} in order to remain as energy efficient as possible.
Usually, a semi-active \emph{WuR} combines an envelope detector that is a passive circuit along with a low-power comparator circuit that is an active component.
It is worth noting that the best obtained performance in semi-active \emph{WuR} systems in terms of sensitivity is observed in \cite{Mango_journal} with a reported sensitivity of \emph{-55 dBm}, while the receiver circuit drains \emph{600 nA} when operating in its quiescent state.
%
The aforementioned circuits have also gain significant attention over the last years with new principles presented in \cite{Nanowatt_clock, proof_wur, ZIPPY, open_wur}.

In this work we leverage the existing semi-active principle and we introduce \emph{eWake}, a novel enhancement that offers substantially increased sensitivity, that goes beyond \emph{-70 dBm}, while the power consumption of our \emph{WuR} circuit remains at the same levels.





\begin{figure*}[!t]
\begin{center}
    \subfigure[\emph{WuR} Architecture]{
      \includegraphics[width=0.38\linewidth]{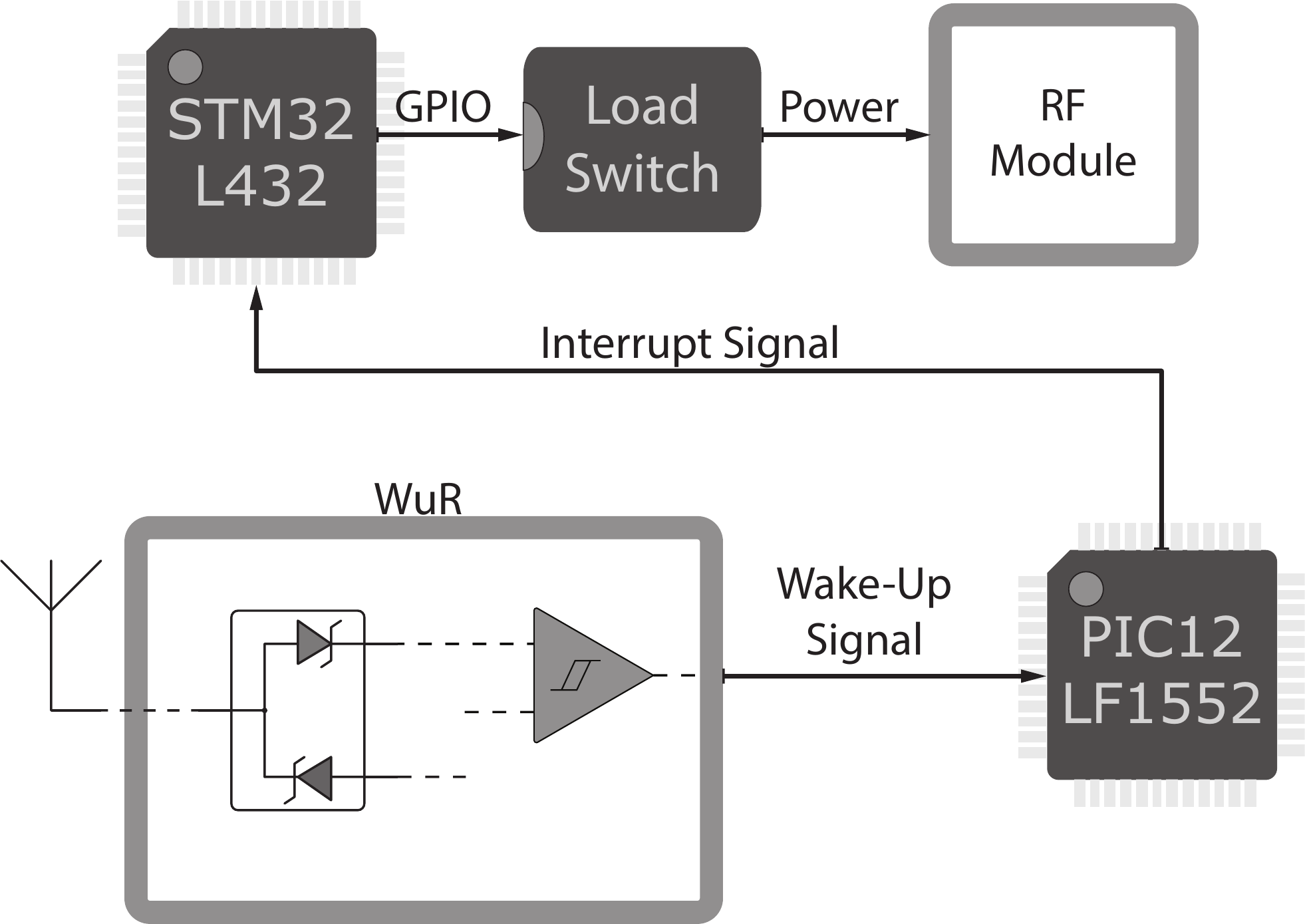}
      \label{fig: genericarch}}
      \hspace{1.0cm}
    \subfigure[The mini ICARUS mote w/ \emph{WuR} Receiver]{
      \includegraphics[width=0.38\linewidth]{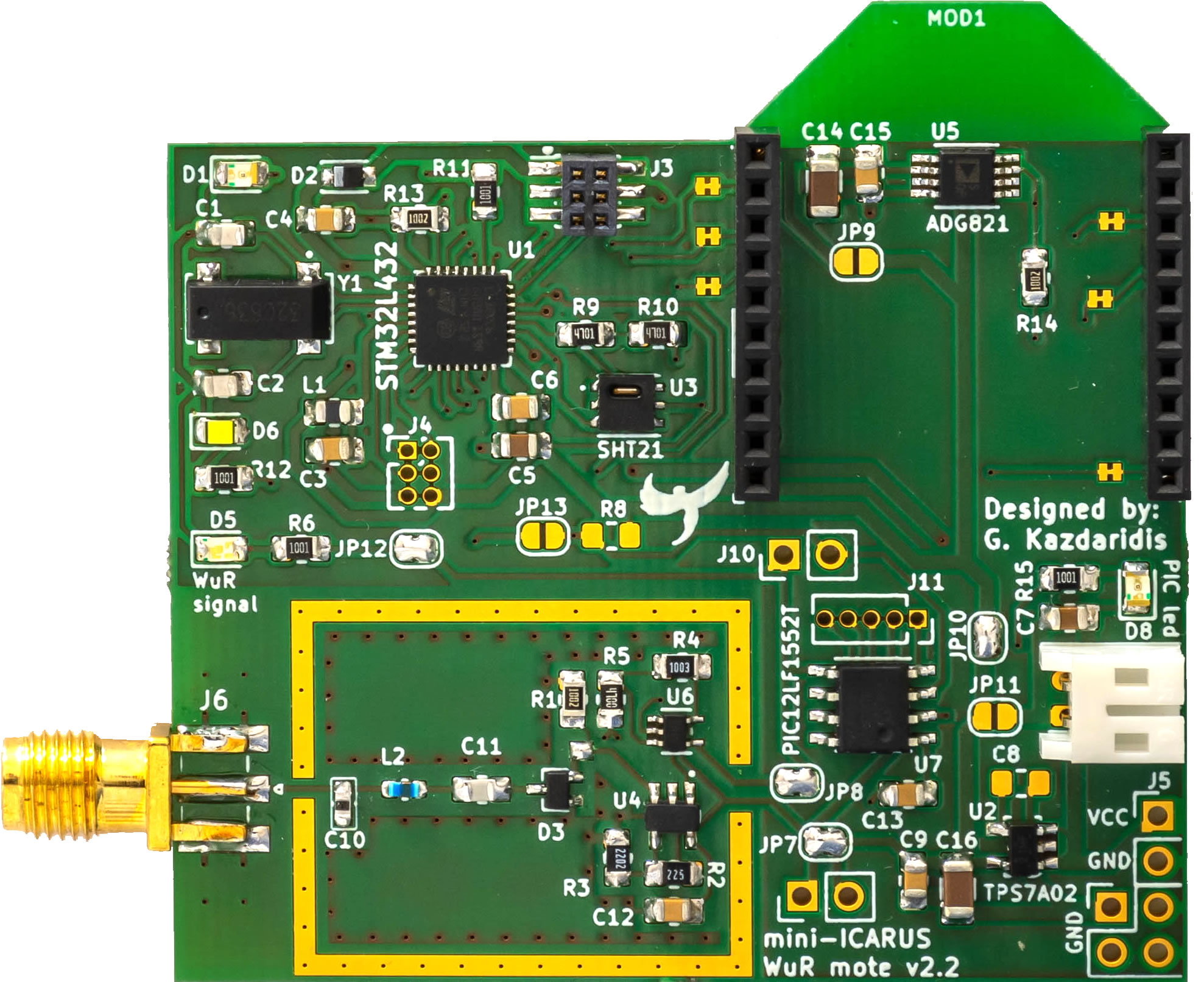}
      \label{fig: mini_icarus}}
      \vspace{-0.45cm}
      \caption{Proposed WuR Architecture \& The mini ICARUS mote w/ the proposed \emph{WuR} Circuit}
           \vspace{-0.3cm}
     \end{center}
\end{figure*}


\vspace{-0.3cm}
\section{Wake-Up System Implementation}\label{system}
\vspace{-0.1cm}

In this section we present the mini-ICARUS mote that integrates a novel WuR circuit able to detect wake-up signals at up to -70 dBm, while we discuss the technical details of the mote and the developed WuR.

\vspace{-0.3cm}
\subsection{System Architecture}\label{arch}
\vspace{-0.1cm}

The architecture of the mini-ICARUS mote is illustrated in the Fig. \ref{fig: genericarch}, while the mote is presented in Fig. \ref{fig: mini_icarus}.
The device integrates a novel \emph{WuR} circuit, an ultra low-power microcontroller (MCU) for supporting addressing capabilities, a host \emph{MCU} and an \emph{XBee footprint} for plugging in wireless interfaces.

The WuR circuit is discussed in detail in next, while the low-power MCU is the \emph{8-bit PIC12LF1552T}. The PIC MCU is configured in its lowest sleep state,
while it is awaken through a raising-edge Interrupt signal upon the reception of a wake-up packet.
The \emph{PIC MCU} process the received signal and identifies whether it should trigger the host \emph{MCU} to wake-up, in case of receiving the correct wake-up packet,
or to return back to the sleep phase if not.
The host \emph{MCU} is the \emph{STM32L432} which is an ultra low-power \emph{ARM Cortex-M4 32-bit RISC MCU} operating at a frequency of up to \emph{80 MHz}.
The \emph{STM32L432} remains in sleep state, consuming only \emph{20 nA}, awaiting from the PIC \emph{MCU} an interrupt signal when a correct wake-up packet is received (containing the proper wake-up address).
Upon the reception of such a packet the \emph{STM32L432} switches to its active state in order to measure, process and transmit data as described by the application scenario.
The mini-ICARUS mote integrates an \emph{XBee footprint} socket where the user can attach the desired wireless interface, such as \emph{IEEE 802.15.4} modules, \emph{LoRa} transceivers or other wireless modems.
Notably, wireless interfaces usually draw significant amount of power even when in sleep \cite{nanothings}, thus we employed a load-switch to entirely turn off the power supply of the wireless module when not in use. The aforementioned method is known as power-gating technique \cite{power_gating}.


\vspace{-0.3cm}
\subsection{Wake-up Receiver Implementation}\label{implementation}
\vspace{-0.1cm}

\begin{figure}[t]
\vspace{-0.0cm}
  \begin{center}
      \includegraphics[width=1.0\columnwidth]{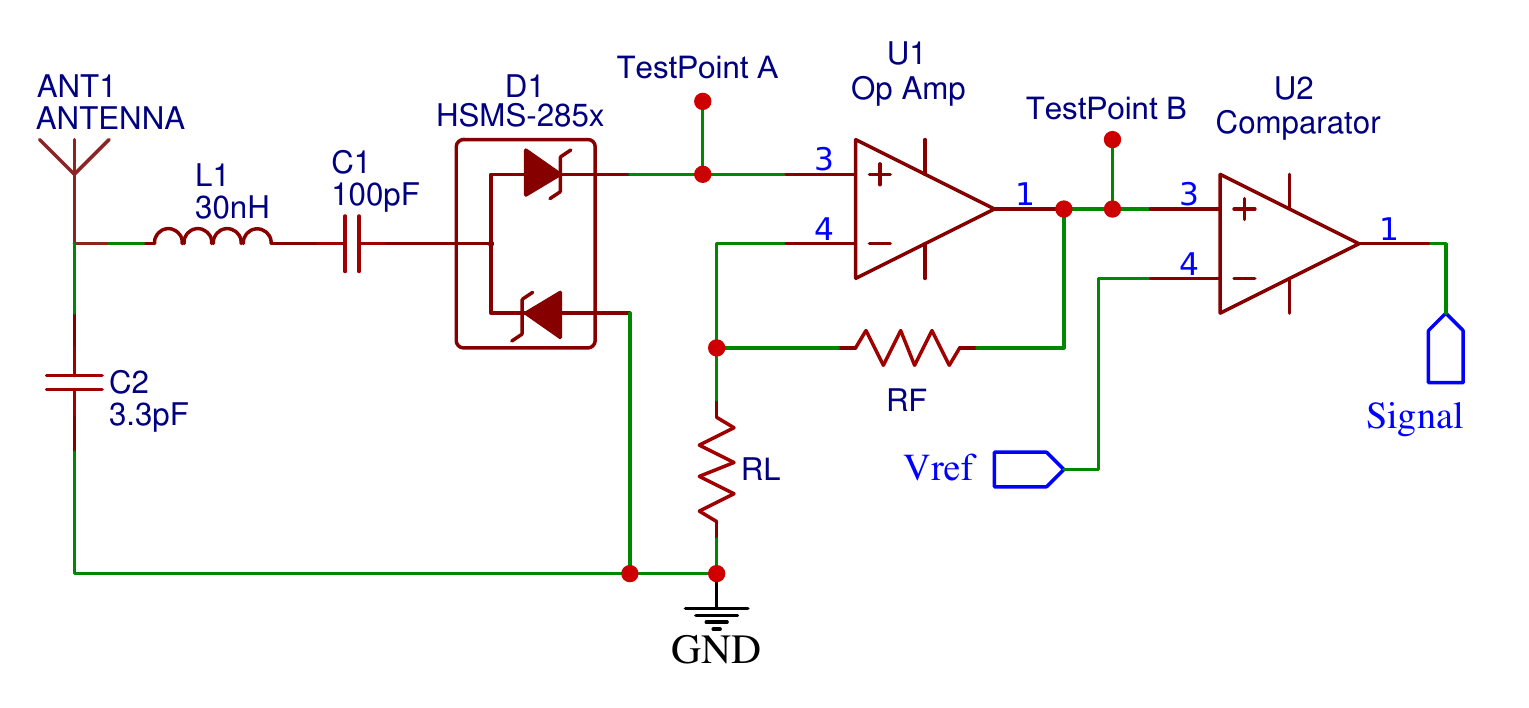}
             \vspace{-0.6cm}
      \caption{Proposed \emph{WuR} Schematic Circuit}
      \vspace{-0.2cm}
  \label{fig: circuit}
  \end{center}
  \vspace{-0.2cm}
  \end{figure}

The developed prototype receiver consists of low-cost off-the-shelf electronics and a low-power micro-controller.
The schematic diagram of the receiver is illustrated in Fig. \ref{fig: circuit}, which is integrated into the mini-ICARUS mote. 
For the wake-up receiver a matching network, a passive rectifier (\emph{HSMS-285c}), an Operation Amplifier and a Comparator \emph{IC} (Integrated Circuit) were used.
Typically, the matching network consists of a capacitor and an inductor element
that together form an L-C network used to match the impedance of the antenna with the rest circuit.
Then, a passive rectifier in the topology of an envelope detector is formed with the aid of two Schottky diodes, used to discard the high frequency signals and to deliver the modulated \textit{OOK} signal.
In our setup we used the HSMS-285c rectifier by Avago Technologies.
Next, the harvested signal is processed by a low-power operation amplifier and a comparator circuit in order to fed it in binary format to the PIC12LF1552T MCU.
The PIC12LF1552T MCU is
responsible for processing the received signal and identifying the acquired address to verify whether it should wake-up the host node or not.
To awake the network's nodes we utilize \emph{868 MHz LoRa} radio transceivers, by modulating the propagated information using \textit{On-Off Keying (OOK)} modulation.

Our finding lies in the observation that the power harvested by the envelope detector (the output signal of the HSMS-285c rectifier - \emph{TestPoint A}, in Fig. \ref{fig: circuit}) is not adequate to trigger the next stage when receiving high attenuated wake-up packets.
%
Notably, in all the previously published works the harvested signal by the envelope detector (\emph{HSMS-285c}) in \emph{TestPoint A} is fed directly to the comparator \emph{IC}, thus the overall sensitivity of the \emph{WuR} is solely depended on the characteristics of the comparator being used.
%
When using a comparator with low Input Offset Voltage  ($V_{OS}$),
as for example the \emph{LPV7215} (300 $\mu$V $V_{OS}$), the \emph{WuR} circuit achieves high sensitivity of \emph{-55 dBm}, while when employing the \emph{TLV3691} (\emph{3 mV} $V_{OS}$) the obtained sensitivity is only at \emph{-32 dBm}.
Of course, the power consumption of these comparators is 
proportional to their performance, meaning that the \emph{TLV3691} draws only \emph{110 nA} while the \emph{LPV7215} consumes roughly \emph{580 nA}.
Table \ref{table:comparators} summarizes the compelling analog comparators along with their characteristics.
Notably, another crucial parameter that plays significant role in the performance of the \emph{WuR} circuit, is the Input Bias Current ($I_{B}$), which is drained by the comparator itself to bias the internal circuitry of the IC.
In our work we propose a new architecture, by first amplifying the harvested RF signal (\emph{TestPoint A}) and then feeding the intensified signal (\emph{TestPoint B}) to a low-power comparator.
Leveraging this strategy we are able to detect RF signals even when receiving highly attenuated packets, thus we achieve sensitivity far beyond the state-of-the-art. Our early experiments have illustrated sensitivity beyond \emph{-70 dBm}.

\begin{figure}[t]
\vspace{-0.0cm}
  \begin{center}
      \includegraphics[width=0.8\columnwidth]{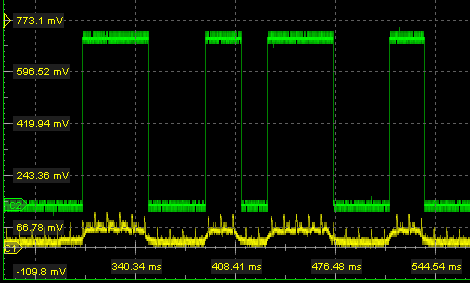}
             \vspace{-0.2cm}
      \caption{Amplified Signal upon Packet Reception at \emph{-60 dBm} (Yellow line) \& Comparator's Output (Green line)}
      \vspace{-0.0cm}
  \label{fig: oscilloscope}
  \end{center}
  \vspace{-0.2cm}
  \end{figure}

Fig. \ref{fig: oscilloscope} illustrates the obtained signal level amplified by the operation amplifier as measured in \emph{TestPoint B} with a yellow line, and the output of the comparator \emph{IC} (labeled as \emph{Signal}) with a green line, when receiving a wake-up packet at \emph{-60 dBm}.

\begin{table}[]
\scalebox{0.95} {
\begin{tabular}{ccccc}
\hline\hline
\textbf{Comparators}    & \textbf{Drain} & \textbf{$V_{OS}$} & \textbf{$I_{B}$} & \textbf{Sensitivity}  \\ \hline\hline
TLV3691                 & 110 nA       & 3 mV             & 80 pA                 & -32 dBm \cite{Mango_journal}    \\ \hline
TLV7031/41              & 335 nA       & 100 $\mu$V           & 2 pA                  & -      \\ \hline
TLV3701                 & 560 nA       & 200 $\mu$V           & 80 pA                 & -      \\ \hline
LPV7215                 & 580 nA       & 300 $\mu$V           & -40 fA                & -55 dBm \cite{Mango_journal}\\ \hline
LTC1540                 & 300 nA       & n/a                  & 10 fA                 & -51 dBm \cite{marinkovic}       \\ \hline
MAX919                  & 350 nA       & 1 mV               & 150 fA                & -      \\ \hline
TS881                   & 260 nA       & 500 $\mu$V           & 1 pA                  & -      \\ \hline
ADCMP380                & 92 nA        & n/a                & 4 nA                    & -      \\ \hline\hline
\end{tabular}
}
\caption{Compelling Comparators and their Specifications}
\vspace{-0.6cm}
\label{table:comparators}
\end{table}

\textbf{Components Selection:}
The Radio Frequency (RF) passive rectifier we used is the HSMS-285c which is the best option for harvesting RF signals,
optimized for frequencies below 1.5 GHz.
As regards the Operation Amplifier we have employed is the \emph{LPV811} that features \emph{450 nA} power draw, configured to amplify the obtained signal a hundredfold.
Next the signal is fed to the comparator \emph{IC}, which in our prototype is the \emph{TLV3691} that consumes roughly \emph{120 nA}.
We opted for the \emph{TLV3691} in spite of featuring high $V_{OS}$, since the proposed circuit is no longer dependent on the comparator's $V_{OS}$.
The overall consumption of \emph{525 nA} can be further optimized by reducing the supply voltage from \emph{3.3V} to \emph{1.6V}.

Regarding the Operation Amplifier and the comparator \emph{ICs}, there are also several other options that could be used instead.
Table \ref{table:opamps} summarizes some recently launched \emph{ICs} by \emph{Texas Instruments} that feature extremely low-power consumption, in the order of a few hundred nA.
Notably, the majority of the general purpose operation amplifiers consume at least a few \emph{$\mu$A}, thus the usage of the proposed amplifiers is deemed necessary.
Apparently, another very attractive option is the \emph{LPV801} amplifier that draws only \emph{320 nA}, while also featuring very low $I_{B}$ of \emph{100 fA}. In future designs we are looking into using the aforementioned amplifier to reduce the overall power consumption. 
By doing so, the overall consumption of the \emph{WuR} circuit will be as low as \emph{450 nA}.
Regarding the employed comparator, is already one of the best options (Table \ref{table:opamps}), since only the \emph{ADCMP380 IC} drains lower power but it comes with a fixed voltage reference of \emph{500 mV} or \emph{1 V}, thus not that applicable in our system. Notalby, currently we use a voltage reference threshold of toughly \emph{27 mV}.

It is worth noting that all the selected components are off-the-shelf and substantially inexpensive, thus the proposed WuR can be easily implemented.


\textbf{Address Matching:}
In order to reduce the overall false positive wake-ups as a result also the power consumption of the entire system,
we employed an ultra low-power \emph{MCU} that process the received signal prior triggering the host \emph{MCU}.
This \emph{MCU} is the \emph{8-bit PIC12LF1552T}.
The \emph{PIC12LF1552T} is configured in sleep state, consuming only 20 nA, while it is awaken upon the reception of a wake-up frame.
Notably, the \emph{PIC12LF1552T} requires a few $\mu$s to wake-up in order to process the signal, thus our wake-up packet contains a preamble with a similar time duration in order to be sure that the \emph{MCU} is ready to read and process the required information.
We selected the PIC \emph{MCU} to provide address matching capabilities, because it drains only 32 $\mu$A per MHz thus considered to be an extremely inexpensive solution in terms of power consumption.
In our application it is configured to operate at \emph{2 MHz}, as also in most related works \cite{Mango_journal}.



\textbf{Matching Network:}
In order to tune the matching network of a \emph{WuR} circuit, a network analyzer is essential since it calculates the exact values required for the \emph{L-C} components employed, to transfer as much power as possible of the received radio signal available on the antenna to the rest of the circuit.
This step is essential, 
since the only way to evaluate the sensitivity of the receiver is to execute realistic experiments with RF transmitters and of course by employing variable attenuators to precisely diminish the transmitted radio signal. 
However, such an apparatus is extremely expensive, therefore not easy to find and use in every laboratory.
In our setup we followed a different approach to tune the matching network, by employing the low-cost \emph{ADALM-PLUTO} \cite{pluto} \emph{Software Defined Radio (SDR)}.
We used the \emph{SDR} platform  to transmit signals in different center frequencies in order to detect the frequency of the utilized \emph{L-C} network that provides less attenuation.
In essence, we plot the performance of the \emph{L-C} network around the given frequency band.
After repeating the same experiment by using different \emph{L-C} values we identified the elements that provide the highest received signal strength in the 868 MHz band.
To characterize the strength of the received signal we measured the harvested voltage output of the \emph{HSMS-285c} rectifier (\emph{TestPoint A} in Fig. \ref{fig: circuit}) of the proposed \emph{WuR} circuit. 
Notably, we easily approached the required \emph{L-C} values by starting our trials with values similar to the ones presented in other works \cite{Akos}.


\begin{table}[]
\scalebox{0.95} {
\begin{tabular}{cccc}
\hline
\textbf{Op. Amp.} & \textbf{Consumption} & \textbf{$V_{OS}$} & \textbf{$I_{B}$} \\ \hline
LPV521          & 350 nA               & 100 $\mu$V           & 40 fA                 \\ \hline
LPV801          & 320 nA               & 550 $\mu$V           & 100 fA                \\ \hline
LPV811          & 450 nA               & 55 $\mu$V            & 100 fA                \\ \hline
LPV821          & 650 nA               & 1.5 $\mu$V           & 7 pA                  \\ \hline
TLV8541         & 480 nA               & 300 $\mu$V           & 100 fA                \\ \hline
TLV8801         & 450 nA               & 550 $\mu$V            & 100 fA                \\ \hline
TLV8811         & 450 nA               & 75 $\mu$V            & 100 fA                \\ \hline
\end{tabular}
}
\caption{Compelling Operation Amplifiers}
\vspace{-0.6cm}
\label{table:opamps}
\end{table}

\textbf{Transmitter Device:}
To awake the network's nodes
we use the \emph{LoRa} \cite{lora} technology, which is a low-power ultra-long range \emph{IoT} technology. 
The wake-up packets are modulated using \textit{OOK} modulation, which in essence means that carrier signal is transmitted to represent the binary one, or suspended to represent the binary zero. Of course, the signal is modulated at a fixed transmission rate, so as the receiver to be able to extract the information by the received signal.
Notably, the modulated packet contains two chunks of information, the \textit{network id} and the \textit{address} of the targeted node. This way, the wake-up circuit, ensures firstly that the packet belongs to its wake-up network and secondly whether the address refers to it.
Of course, sensor nodes, can support more than one wake-up addresses, so that we can awaken a set of nodes at the same time.
It is worth noting that the wake-up signal begins with an artificial delay 
to allow the activation of the \emph{PIC12LF1552T MCU}, prior to the reception of the modulated information.


\vspace{-0.2cm}
\section{Conclusions}\label{conclusion}
\vspace{-0.1cm}
In this work we showcase a new scheme for semi-passive \emph{WuR} circuits that remarkably increases the sensitivity of the 
existing state-of-the-art implementations, reaching beyond \emph{-70 dBm}.
Our finding lies in the employment of a nano-power amplifier that intensifies the signal prior to the decoding process.
The overall consumption of our proposed circuit is roughly \emph{580 nA}, while it can be substantially reduced by using alternative \emph{ICs}.
Lastly, our future plans include the measurement and characterization of the proposed \emph{WuR's} power consumption profile  using our power monitoring tools \cite{eProfiler, insitu}.



\balance

\bibliographystyle{ACM-Reference-Format}
\bibliography{sample-base}

\end{document}